\documentclass[prd,twocolumn,showpacs,amsmath,amssymb,citeautoscript,nobibnotes,preprintnumbers]{revtex4-1}

\usepackage{graphicx,times}
\usepackage{array}
\usepackage{bm}
\usepackage{amsfonts}

\newcommand{\beq}{\begin{equation}}
\newcommand{\eeq}{\end{equation}}
\newcommand{\beqa}{\begin{eqnarray}}
\newcommand{\eeqa}{\end{eqnarray}}

\newcommand{\psib}{{\overline{\psi}}}

\newcommand\comment[1]{}

\begin{document}

\title{Solutions to sign problems in lattice Yukawa models}
\author{Shailesh Chandrasekharan}
\affiliation{Department of Physics, Duke University, Durham, NC 27708, USA}

\begin{abstract}
We prove that sign problems in the traditional approach to some lattice Yukawa models can be completely solved when fermions are formulated using fermion bags and bosons are formulated in the worldline representation. We prove this within the context of two examples of three dimensional models, symmetric under $U_L(1) \times U_R(1) \times Z_2 (\mbox{Parity})$ transformations, one involving staggered fermions and the other involving Wilson fermions. We argue that these models have interesting quantum phase transitions that can now be studied using Monte Carlo methods.
\end{abstract}
\maketitle

\section{Introduction}

Feynman path integrals can be used to map a quantum statistical mechanics partition function into  a classical statistical mechanics partition function with one caveat : the Boltzmann weight of the classical partition function may be negative or even complex. When this occurs the mapping is said to suffer from a sign problem since the mapping is not useful for Monte Carlo methods. However, fortunately the mapping is not unique and it may be possible to find a different mapping in which the Boltzmann weights of the classical partition function are indeed positive. If these weights are calculable with polynomial effort as the system size grows, the mapping is said to be free from a sign problem and one may be able to construct a Monte Carlo algorithm to sample the classical configuration space and solve the problem. In many interesting cases a mapping without a sign problem has eluded researchers and discovering the correct mapping is defined as a solution to the sign problem.

Sign problems are well known obstacles to solving many quantum field theories from first principles. This is particularly true when the microscopic degrees of freedom contain fermions. Famous examples of physical systems where sign problems have hindered progress are a finite density of strongly interacting matter \cite{deForcrand:2010ys,Gupta:2011ma,Levkova:2012jd} and a finite density of electrons with Coulomb repulsion \cite{PhysRevB.41.9301}. Many other model field theories also suffer from sign problems.  Examples of these include four-fermion field theories and interacting boson-fermion (Yukawa) models. These often have interesting low energy physics and are used as effective field theories \cite{Rosenstein:1990nm,Epelbaum:2009pd}. In three dimensions they can contain interesting quantum phase transitions that have remained unexplored due to sign problems. 

While some important sign problems may be unsolvable \cite{Troyer:2004ge}, it has become clear that a variety of sign problems can indeed be solved by finding the right representation of the partition function. Novel solutions to sign problems have been found in both purely bosonic models with a complex action \cite{Bietenholz:1995zk,Alford:2001ug,PhysRevD.75.065012,PhysRevD.81.125007,Bloch:2011jx,Gattringer:2011gq} and in purely fermionic models \cite{PhysRevLett.83.3116,Chandrasekharan:2012va}. Here, for the first time we show that sign problems in models containing fermions and bosons as dynamical fields interacting with each other (which we call Yukawa models) can also be solved. We illustrate this using two examples of lattice models in three dimensions.

In section 2 we review the sign problems that haunt the traditional approach to two lattice Yukawa models one with staggered fermions and one with Wilson fermions. In section 3 we show how the fermion-bag approach when combined with the worldline formulation of bosonic degrees of freedom solves these sign problems. In section 4 we present our conclusions.

\section{Sign Problems}

\begin{figure}
\includegraphics[width=0.4\textwidth]{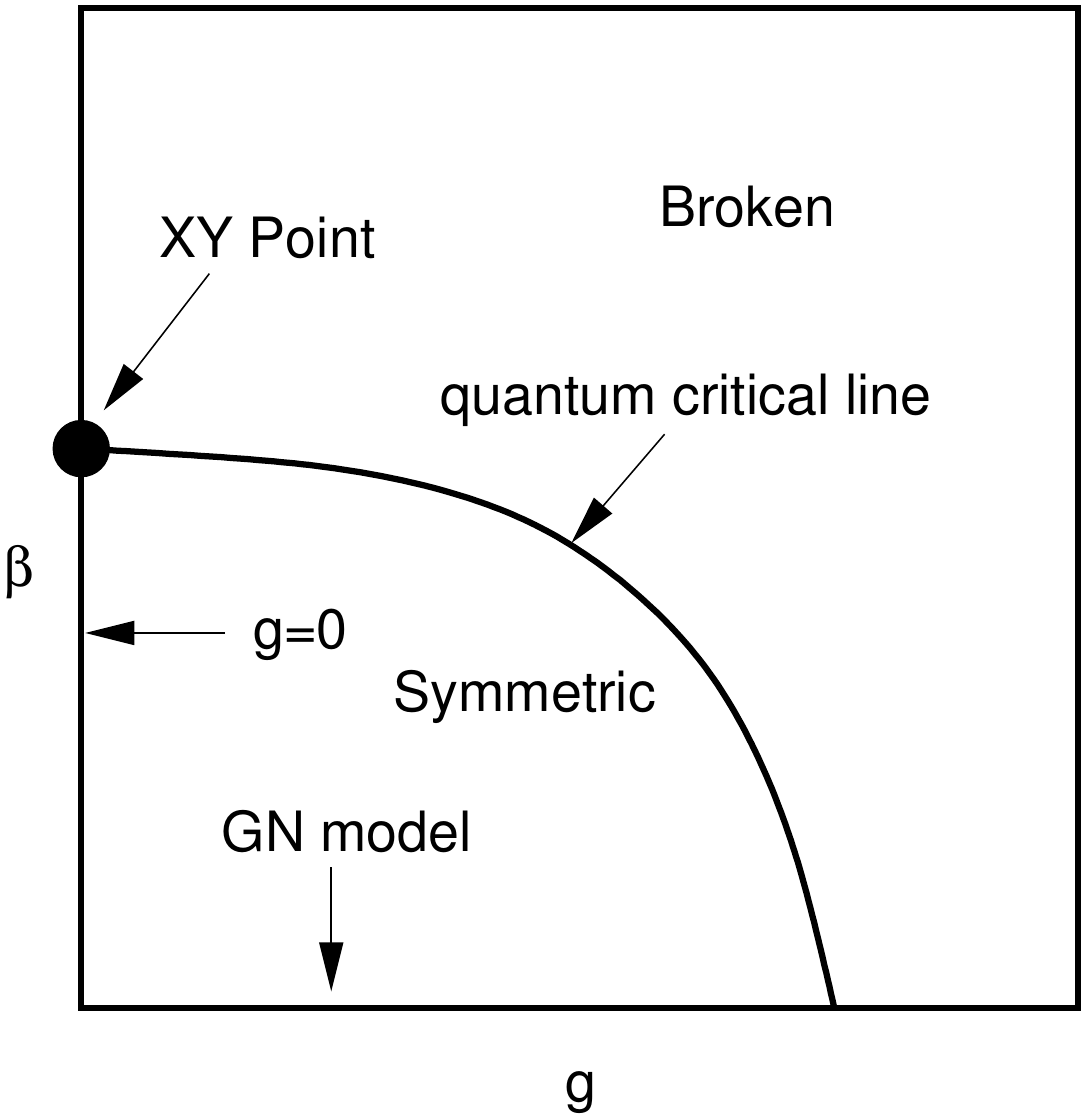}
\caption{A schematic phase diagram of $U_L(1)\times U_R(1) \times Z_2$ symmetric lattice Yukawa models discussed in the text. The symmetric phase contains massless fermions while the broken phase contains massless bosons. The critical point on the $g=0$ line is the $XY$ critical point. For small $\beta$ the model reduces to a Gross-Neveu (GN) model.}
\label{fig1}
\end{figure}

In this section we review two examples of sign problems in lattice Yukawa models, one with staggered fermions and the other with Wilson fermions. For convenience our models are defined in three space-time dimensions, but the discussion is applicable in higher dimensions with minor modifications. We begin with the example with staggered fermions. The action of the model is given by 
\begin{equation}
S_s = \sum_{x,y} \ \psib_x \ (D^s[\theta])_{xy} \ \psi_y + S_b[\theta]
\label{stact}
\end{equation}
where $\psib_x,\psi_x$ are two Grassmann valued fields on the lattice site $x \equiv (x_1,x_2,x_3)$ of a cubic lattice with $V$ sites. While there are many choices for the bosonic action, for simplicity in this work we choose it to be the classical $XY$ model,
\begin{equation}
S_b[\theta] = - \beta \sum_{\langle xy\rangle} \cos(\theta_x - \theta_y).
\end{equation}
where the bosonic field $\theta_x$ is a phase. Here $\langle xy\rangle$ refers to nearest neighbor sites. The matrix $D^s[\theta]$ is the $V \times V$ staggered Dirac operator whose matrix elements are given by
\begin{subequations}
\begin{eqnarray}
(D^s[\theta])_{xy} &=& -g \ \mathrm{e}^{i\varepsilon_x \theta_x} \delta_{x,y} + (D^{s_0})_{xy},
\label{stagg1}
\\
 (D^{s_0})_{xy} &=& \sum_\alpha \eta_{\alpha,x} \nabla^\alpha_{xy},
\label{stagg2}
\end{eqnarray}
\end{subequations}
where the fluctuating mass term depends on the bosonic field. The index $\alpha=1,2,3$ represents the three directions, $\eta_{\alpha,x}$ are the staggered fermion phase factors ($\eta_1 = 1,\eta_2 = (-1)^{x_1}, \eta_3= (-1)^{x_1+x_2}$), $\varepsilon_x = (-1)^{x_1+x_2+x_3}$ is the site parity and
\begin{equation}
\nabla^\alpha_{xy} = \frac{1}{2} (\delta_{x,y+\hat{\alpha}} - \delta_{x+\hat{\alpha},y}).
\label{nabla}
\end{equation}
The parameters $\beta$ and $g$ control the physics of the model. It is easy to verify that the action in Eq.~(\ref{stact}) is invariant under the following $U_L(1) \times U_R(1)$ chiral transformations :
\begin{subequations}
\begin{eqnarray}
\psi_x &\rightarrow& \mathrm{e}^{i\theta_L (1+\varepsilon_x)/2+i\theta_R(1-\varepsilon_x)/2}\ \psi_x,
\\
\psib_x &\rightarrow& \mathrm{e}^{-i\theta_L (1-\varepsilon_x)/2-i\theta_R(1+\varepsilon_x)/2}\ \psib_x, 
\\
\theta_x &\rightarrow& \theta_x + (\theta_R - \theta_L)\varepsilon_x
\end{eqnarray}
\end{subequations}
In addition, the action is also invariant under the $Z_2$ (parity) transformations:
\begin{equation}
x \rightarrow -x,\ \ \psib_x \rightarrow -\psib_{-x},\ \ \psi_x \rightarrow \psi_{-x},\ \ \theta_x \rightarrow \theta_{-x} + \pi
\end{equation}
These symmetries play an important role in governing the long distance physics of the model. The model possesses at least two phases, a symmetric phase with massless fermions and massive bosons and a broken phase with massless bosons and a massive fermion. Since a fermion mass term breaks parity in three dimensions, it is natural to expect massive fermions in the broken phase. Both the phases can be accessed by tuning $\beta$ and $g$. A schematic phase diagram, expected from general arguments, is shown in Fig.~\ref{fig1}. The critical point on the $g=0$ axis is the well known 3d-$XY$ critical point. If it exists, a second order quantum critical line separating the two phases starting from the $XY$ point must be governed by a different critical point. Based on symmetries we conjecture that it belongs to the universality class of the model studied in \cite{PhysRevLett.108.140404}. Unfortunately, as we will discuss below, this lattice Yukawa model cannot be studied using traditional Monte Carlo methods due to sign problems.

Next we consider a Yukawa model constructed with Wilson fermions. We again restrict ourselves to three space-time dimensions for simplicity. The action of the model is given by
\begin{equation}
S_w = \sum_{xy} \psib_x \ (D^w[\theta])_{xy} \ \psi_y + S_b[\theta]
\label{wact}
\end{equation}
where, unlike the staggered fermion case, the fields $\psib_x$ and $\psi_x$ are each four component fields, written in terms of four two-component left and right fields as
\begin{equation}
\psib = \left(\begin{array}{cc} \psib^L & \psib^R\end{array}\right),\ \ 
\psi = \left(\begin{array}{c} \psi^L \cr \psi^R\end{array}\right).
\end{equation}
We label the eight Grassmann fields on each site as $\psib^c_{s,x},\psi^c_{s,x}$ where $c=L,R$ and $s=1,2$. The Dirac operator is given by
\begin{equation}
(D^w[\theta])_{xy} = \left(\begin{array}{cc} 
(D^{w_0})_{xy} & -g \ \mathbf{I}\ \mathrm{e}^{-i\theta_x} \delta_{xy} \cr
g\ \mathbf{I}\ \mathrm{e}^{i\theta_x} \delta_{xy} & ({D^{w_0}}^\dagger)_{xy} \end{array}\right),
\end{equation}
where $\vec{\sigma}$ are the Pauli matrices and $\mathbf{I}$ is the $2 \times 2$ identity matrix. The $2V \times 2V$ matrix $D^{w_0}$ is the Wilson-Dirac operator defined by
\begin{equation}
(D^{w_0})_{xy} =  \mathbf{I} \ M_{xy}\ + \vec{\sigma}\cdot \vec{\nabla}_{xy}
\end{equation}
where $\vec{\nabla}$ was defined in Eq.~(\ref{nabla}) and $M$ is the $V \times V$ Wilson mass matrix defined by
\begin{equation}
M_{xy} =  - \frac{1}{\kappa} \delta_{x,y}   + \frac{1}{2} \sum_\alpha (\delta_{x,y+\hat{\alpha}} + \delta_{x+\hat{\alpha},y}).
\end{equation}
The action (\ref{wact}) is also invariant under the following $U_L(1) \times U_R(1)$ transformations:
\begin{subequations}
\begin{eqnarray}
&&\psi^L_x \rightarrow \mathrm{e}^{i\theta_L}\psi^L_x,\  \psib^L_x \rightarrow \mathrm{e}^{-i\theta_L}\psib^L_x,
\\
&& \psi^R_x \rightarrow \mathrm{e}^{i\theta_R}\psi^R_x,\ \psib^R_x \rightarrow \mathrm{e}^{-i\theta_R}\psib^R_x,
\\
&&\theta_x \rightarrow \theta_x + (\theta_R - \theta_L).
\end{eqnarray}
\end{subequations}
However, due to the presence of the Wilson mass term $M_{xy}$, the action is not invariant under the $Z_2$ (parity) transformations
\begin{equation}
x \rightarrow -x,\ \ \psib_x \rightarrow -\psib_{-x},\ \ \psi_x \rightarrow \psi_{-x},\ \ \theta_x \rightarrow \theta_{-x} + \pi.
\end{equation}
Parity transformations are restored in the long distance physics, when the hopping parameter $\kappa$ is tuned to a critical value where massless fermions emerge. For $g=0$, one finds $\kappa_c = 1/3$. In general $\kappa_c$ is a function of the couplings $g$ and $\beta$. Once $\kappa$ is tuned to this value the phase diagram of the model as a function of $\beta$ and $g$ is very similar to Fig.~\ref{fig1}, except that the critical behavior on the quantum critical line could belong to a different universality class as compared to the staggered fermion case. The difference is essentially due to the number of fermions that become massless on the critical line.

In order to study the two Yukawa models discussed above one begins with the partition function
\begin{equation}
Z_i = \int [d\psib\ d\psi] \ \mathrm{e}^{-S_i},
\end{equation}
where $i=s,w$ and tries to construct a Monte Carlo technique to compute expectation values of appropriate quantities. The traditional method is to integrate the fermions completely and rewrite
\begin{equation}
Z_i = \int [d\theta] \ \mathrm{e}^{-S_b[\theta]} \ \mathrm{Det}(D^i[\theta]).
\end{equation}
If the determinant of the Dirac operator $D^i[\theta]$ was non-negative then one could have devised a Monte Carlo method to sample the $[\theta]$ configurations. Unfortunately this is not the case for both the models considered here. With staggered fermions there is no symmetry that can be used to even argue that $\mathrm{Det}(D^s[\theta])$ is real. The determinant can in fact be complex. On the other hand with Wilson fermions one can prove that
\begin{equation}
C(D^w[\theta])C = (D^w[\theta])^*, \ \mbox{where}\ \ 
C = \left(\begin{array}{cc} 0 & -i\sigma_2 \cr i \sigma_2 & 0 \end{array}\right),
\end{equation}
which means the determinant of $D^w[\theta]$ is real. Unfortunately, it could still be negative. Thus, both Yukawa models discussed above suffer from sign problems in the traditional method. As far as we know, these sign problems have remained unsolved until now. While these sign problems clearly arise for reasons completely different from the sign problem in finite density QCD \cite{Levkova:2012jd}, they seem equally difficult.

\section{Fermion-Bag Worldline Approach}

We now prove that the sign problems disappear in the above two models when the fermions are formulated in the fermion-bag approach and bosons are formulated in the worldline representation. Since the details of the proof are slightly different in each case, we will discuss them separately.  Let us first consider the partition function of the Yukawa model with staggered fermions. As explained in \cite{Chandrasekharan:2009wc,PhysRevLett.108.140404} we expand the interaction in powers of the coupling. We begin by noting that
\begin{equation}
\mathrm{e}^{g (\psib_x \psi_x)\mathrm{e}^{i\varepsilon_x\theta_x}} = 1 + g (\psib_x \psi_x) \mathrm{e}^{i\varepsilon_x\theta_x},
\end{equation}
due to the Grassmann nature of $\psib_x\psi_x$. We can represent the two terms on the right through the discrete variables $n_x = 0,1$. The first term refers to $n_x = 0$ or no interaction while the second term represents $n_x = 1$ and indicates the presence of an interaction vertex $\psib_x\psi_x$ (or a monomer). Every monomer is also associated with the term $\mathrm{e}^{i\varepsilon_x\theta_x}$. Using this idea it is possible to write
\begin{eqnarray}
Z_s &=& \sum_{[n]} \ g^j \Big\{ \int [d\psib\ d\psi] \ \mathrm{e}^{-S^s_0 } \ \psib_{x_1}\psi_{x_1}... \psib_{x_j}\psi_{x_j}
\nonumber \\
&\times&  \int [d\theta] \ \mathrm{e}^{-S_b[\theta]}\  \mathrm{e}^{i\varepsilon_{x_1}\theta_{x_1}} ... \mathrm{e}^{i\varepsilon_{x_j}\theta_{x_j}} \Big\}
\label{fbag1}
\end{eqnarray}
where $[n]$ represents a configuration of monomers, $j$ refers to the total number of monomers and 
\begin{equation}
S^s_0 = \sum_{x,y} \psib_x\ (D^{s_0})_{xy}\ \psi_y
\end{equation}
is the free staggered fermion action. . For every configuration $[n]$ we have labeled the sites where the monomers are located as $\{x_1,...,x_j\}$.  As explained in \cite{Chandrasekharan:2009wc,Chandrasekharan:2011vy} it is possible to show that
\begin{equation}
\int [d\psib\ d\psi] \ \mathrm{e}^{-S^s_0 } \ \psib_{x_1}\psi_{x_1}... \psib_{x_j}\psi_{x_j} = \mathrm{Det}(W[n]) \geq 0
\end{equation}
where $W[n]$ is a $(V-j) \times (V-j)$ matrix which has the same elements as $D^{s_0}$ except that the sites $x_1,..,x_j$ are dropped. The sites $(V-j)$ form what we call the strong coupling fermion bag \cite{Chandrasekharan:2011vy,PhysRevLett.108.140404}. Interestingly, one can also rewrite the bosonic integral using the worldline representation as discussed in \cite{Chandrasekharan:2008gp,PhysRevD.81.125007}. One finds that
\begin{eqnarray}
&&  \int [d\theta] \ \mathrm{e}^{-S_b[\theta]}\  \mathrm{e}^{i\varepsilon_{x_1}\theta_{x_1}} ... \mathrm{e}^{i\varepsilon_{x_j}\theta_{x_j}} =
\sum_{[k]} \prod_{x,\alpha} \ \{I_{k_{x,\alpha}}(\beta)\}
\nonumber \\
&& \ \ \ \ \ \ \times \ \prod_x \delta\Big(\sum_\alpha(k_{x,\alpha} - k_{x-\alpha,\alpha}) + \varepsilon_x n_x\Big)
\label{worldline}
\end{eqnarray}
where the integer bond variables $k_{x,\alpha}$ represent worldlines of charged particles, $I_{k_{x,\alpha}}(\beta)$ is the modified Bessel function, $[k]$ represents a configuration of these worldlines. Note that the interaction between the fermions and bosons appear through the constrained delta function which essentially implies that every monomer either creates or destroys bosonic particles. Thus, using a combined fermion-bag-worldline representation, the partition function of the staggered Yukawa model takes the form
\begin{eqnarray}
Z_s &=& \sum_{[n,k]}\ g^j \ \mathrm{Det}(W[n])  \ \prod_{x,\alpha} \ \{I_{k_{x,\alpha}} (\beta)\} 
\nonumber \\
&& \times \ \prod_x \delta \Big(\sum_\alpha(k_{x,\alpha} - k_{x-\alpha,\alpha}) + \varepsilon_x n_x\Big)
\end{eqnarray}
The Boltzmann weight of every $[n,k]$ configuration is non-negative and the sign problem is absent.

Let us now turn to the lattice Yukawa model with Wilson fermions. Unlike the staggered case, we now have four couplings $g\psib^L_{1,x} \psi^R_{1,x} \mathrm{e}^{-i\theta_x}, g\psib^L_{2,x} \psi^R_{2,x} \mathrm{e}^{-i\theta_x}, -g\psib^R_{1,x} \psi^L_{1,x}\mathrm{e}^{i\theta_x}$, and $-g\psib^R_{2,x} \psi^L_{2,x}\mathrm{e}^{i\theta_x}$. We expand in each of these couplings and write
\begin{equation}
\mathrm{e}^{g\ \psib^L_{1,x}\psi^R_{1,x}\ \mathrm{e}^{i\theta_x}} = 1 + g\ \psib^L_{1,x}\psi^R_{1,x} \ \mathrm{e}^{i\theta_x}
\end{equation}
for each of the four couplings. We then introduce four types of monomers $n_{1,x},n_{2,x},n_{3,x},n_{4,x} = 0,1$ at each site $x$ representing these four couplings. For every monomer configuration $[n]$ we label the sites where $n_{1,x}, n_{2,x}, n_{3,x}$ and $n_{4,x}$ are non-zero, as $w_1,...,w_{j_1}$, $x_1,...,x_{j_2}$, $y_1,...,y_{j_3}$, and $z_1,...,z_{j_4}$ respectively. Thus, in the fermion-bag approach we can write
\begin{eqnarray}
Z_w &=& \sum_{[n]} \ g^{j_1+j_2+j_3+j_4} \Bigg\{ \int [d\psib\ d\psi] \ \mathrm{e}^{-S^w_0 } (-1)^{j_3+j_4}
\nonumber \\
&\times &\ \psib^L_{1,w_1}\psi^R_{1,w_1}... \psib^L_{1,w_{j_1}}\psi^R_{1,w_{j_1}} \ 
 \psib^L_{2,x_1}\psi^R_{2,x_1}... \psib^L_{2,x_{j_2}}\psi^R_{2,x_{j_2}}
\nonumber \\
&\times&   \psib^R_{1,y_1}\psi^L_{1,y_1}... \psib^R_{1,y_{j_3}}\psi^L_{1,y_{j_3}}\ 
\psib^R_{2,z_1}\psi^L_{2,z_1}... \psib^R_{2,z_{j_4}}\psi^L_{2,z_{j_4}}
\nonumber \\
&\times&  \int [d\theta] \ \mathrm{e}^{-S_b[\theta]}\  \mathrm{e}^{-i\theta_{w_1}} ... \mathrm{e}^{-i\theta_{x_1}} ... 
\mathrm{e}^{i\theta_{y_1}} ... \mathrm{e}^{i\theta_{z_1}} ...  \Bigg\}
\label{fbag2}
\end{eqnarray}
where 
\begin{equation}
S^w_0 = \sum_{x,y} \Big\{\ \psib^L_x\ (D^{w_0})_{xy}\ \psi^L_y + \psib^R_x\ ({D^{w_0}}^\dagger)_{xy} \ \psi^R_y\Big\}
\end{equation}
is the free Wilson fermion action. Based on the fermion integral it is easy to see that only those monomer configurations which satisfy $j \equiv  j_1+j_2 =j_3+j_4$ contribute to the path integral. If we define the free fermion propagator 
\begin{equation}
G_{s,x;s',y} = \int [d\psib\ d\psi] \ \mathrm{e}^{-S^w_0 } \psi^L_{s,x} \psib^L_{s',y}  
\label{fprop}
\end{equation}
then it is easy to prove that
\begin{equation}
(G_{s,x;s',y})^* = \int [d\psib\ d\psi] \ \mathrm{e}^{-S^w_0 } \ (-1) \ \psib^R_{s,x} \psi^R_{s',y}  
\end{equation}
Using this result along with Wick's theorem it is possible to show that
\begin{eqnarray}
 && \int [d\psib\ d\psi] \ \mathrm{e}^{-S^w_0 } (-1)^{j_3+j_4} \ \psib^L_{1,w_1}\psi^R_{1,w_1}...\psib^L_{1,w_{j_1}}\psi^R_{1,w_{j_1}} 
\nonumber \\
&& \times \ \psib^L_{2,x_1}\psi^R_{2,x_1}... \psib^L_{2,x_{j_2}}\psi^R_{2,x_{j_2}} \psib^R_{1,y_1}\psi^L_{1,y_1}... \psib^R_{1,y_{j_3}}\psi^L_{1,y_{j_3}}
\nonumber \\
&& \times \ \psib^R_{2,z_1}\psi^L_{2,z_1}... \psib^R_{2,z_{j_4}}\psi^L_{2,z_{j_4}} \ =\  |\mathrm{Det}(G[n])|^2
\end{eqnarray}
where $G([n])$ is a $j \times j$ matrix whose matrix elements are the free fermion propagators defined in Eq.~(\ref{fprop}) from $(s=1,\{w_1,...x_{j_1}\})$ and $(s=2,\{x_1,...,x_{j_2}\})$ to $(s=1,\{y_1,...y_{j_3}\})$ and $(s=2,\{z_1,...,z_{j_4}\})$. Combining this result with the worldline representation of the bosonic integral, (similar to Eq.~(\ref{worldline})), we see that
\begin{eqnarray}
&& Z_w = \sum_{[n,k]} \ g^{2j}  \ |\mathrm{Det}(G[n])|^2\ \prod_{x,\alpha} \ \{I_{k_{x,\alpha}} (\beta)\} \ \times
\nonumber \\
&& \prod_x \delta\Big(\sum_\alpha(k_{x,\alpha} - k_{x-\alpha,\alpha}) -  n_{1,x} - n_{2,x} + n_{3,x} + n_{4,x})\Big)
 \nonumber \\
\end{eqnarray}
The above expansion of the partition function is again free of any sign problem since the Boltzmann weights are non-negative.

\section{Discussion}

In this work we have shown that the fermion-bag approach along with the worldline representation of bosonic degrees of freedom, allows us to solve sign problems in some lattice Yukawa models. While the solutions presented here depend on the details of the models considered, the idea is more general and can be applied to solve other sign problems, including those in non-relativistic field theories. For example, models that contain pairing interactions of the form $g \psi^L_s\psi^R_s \mathrm{e}^{i\theta}$ can also be solved. Further, we can study models with discrete rather than continuous symmetries. For example, if we  replace Eq.~(\ref{stagg1}) by
\begin{equation}
(D^s[\theta])_{xy} = -g \ \sin(\theta_x) \delta_{x,y} + (D^{s_0})_{xy},
\end{equation}
the $U_L(1) \times U_R(1)$ symmetry of the model is reduced to a $U_f(1) \times Z_2$. While the conventional approach still suffers from a sign problem, the fermion bag approach does not. Sign problems can also be solved with $N_f$ fermion flavors as long as all fermions couple to boson fields as in the one flavor case. This enhances the symmetry of the models by an $SU(N_f)$ factor.

The bosonic action can also be changed. For example instead of the $XY$ model action, one can also choose the more standard form where
\begin{equation}
S_b[\phi] = -
\sum_{x,\alpha} (\phi^*_x \phi_{x+\alpha} + \phi^*_{x+\alpha} \phi_x)
+ \sum_x \Big( \mu |\phi_x|^2 + \lambda |\phi_x|^4\Big).
\end{equation}
In this case the phase $\mathrm{e}^{i\theta_x}$ in the couplings is replaced by a the complex field $\phi_x$ itself. Then the bosonic integrals in Eqs.~(\ref{fbag1}) and (\ref{fbag2}) can again be represented in a worldline representation with positive Boltzmann weights. Indeed, by writing $\phi_x = \rho_x \mathrm{e}^{i\theta_x}$ in the polar form one can show that
\begin{eqnarray}
&&  \int [d\phi] \ \mathrm{e}^{-S_b[\phi]}\ \phi_{w_1} ...\phi_{w_j} \phi^*_{z_1} ...\phi^*_{z_j}
\nonumber \\
&& = \ \int [d\rho] \ \mathrm{e}^{-\sum_x (\mu \rho^2_x + \lambda \rho_x^4)}
\ \rho_{w_1}..\rho_{w_j} \rho_{z_1}..\rho_{z_j}\ \ 
\nonumber \\
&& \times \Bigg\{ \sum_{[k]} \ \ \prod_{x,\alpha} \ \{I_{k_{x,\alpha}}(2\rho_x\rho_{x+\alpha})\}
\nonumber \\
&& \hspace*{0.5in}
\times \ \prod_x \delta\Big(\sum_\alpha(k_{x,\alpha} - k_{x-\alpha,\alpha}) + n_x\Big)
\Bigg\}
\label{worldline}
\end{eqnarray}
where $n_x = 1$ for $x= w_1,..,w_j$ and $n_x = -1$ for $x = z_1,..,z_j$ and $n_x = 0$ otherwise. Thus, the bosonic integral can again be represented without a sign problem.

An important lesson from our work is that solutions to sign problems only emerge when along with pairing in the fermionic sector, the correct bosonic variables can also be identified. Sign problems in theories with multicomponent boson fields continue to remain a challenge, although our ideas may provide hints for finding a solution. In certain cases we believe that positivity may emerge in terms of quantities like fermionants which unfortunately can be exponentially difficult to compute \cite{Chandrasekharan:2011an}. The ability to convert a quantum partition function into a classical partition function should ultimately be dependent on the underlying physics and the degrees of freedom that capture it optimally.

\section*{Acknowledgments}

I would like to thank H. Baranger, Ph. de Forcrand, M. Hastings, D. Kaplan, A. Li and U.-J. Wiese for helpful discussions. This work was supported in part by the Department of Energy grant DE-FG02-05ER41368.

\bibliography{ref}
\end{document}